# Protein Function Prediction Based on Kernel Logistic Regression with 2-order Graphic Neighbor Information

Jingwei Liu [1]

School of Mathematics and System Sciences,Beihang University,Beijing, P.R China, 100191.

July 19, 2012

## Abstract

To enhance the accuracy of protein–protein interaction function prediction, a 2-order graphic neighbor information feature extraction method based on undirected simple graph is proposed in this paper , which extends the 1-order graphic neighbor featureextraction method. And the chi-square test statistical method is also involved in feature combination. To demonstrate the effectiveness of our 2-order graphic neighbor feature, four logistic regression models (logistic regression (*abbrev.* LR), diffusion kernel logistic regression (*abbrev.* DKLR), polynomial kernel logistic regression (*abbrev.* PKLR), and radial basis function (RBF) based kernel logistic regression (*abbrev.* RBF KLR)) are investigated on the two feature sets. The experimental results of protein function prediction of Yeast Proteome Database (YPD) using the the protein-protein interaction data of Munich Information Center for Protein Sequences (MIPS) show that 2-order graphic neighbor information of proteins can significantly improve the average overall percentage of protein function prediction especially with RBF KLR. And, with a new 5-top chi-square feature combination method, RBF KLR can achieve 99.05% average overall percentage on 2-order neighbor feature combination set.

## 1. Introduction

Proteins are composed of sequences of amino acids and participate in nearly every vital process within cells. With the development of gene chip technique, the interaction of proteins can be correctly measured out, however, many of the functions of proteins are still unknown. Predicting the protein function helps not only to discover the protein's unknown function, but also to detect the organism disease. ( Shalgi, *st al.* 2007; Varnholt *st al.* 2008; Hu, *st al.* 2008 ) As novel feature extraction and classifiers development are two key techniques in protein–protein interaction (PPI) function prediction, we propose a 2-order graphic neighbor feature extraction method based on graph theory and apply the RBF kernel logistic regression to protein–protein interaction prediction in this paper.
To extract the feature of PPI data and predict the protein function, Schwikowski, *et al.*

---

[1]Corresponding author. jwliu@buaa.edu.cn (J.W Liu)

(2000) proposed to assign the possible functions to a protein based on the known functions of its interacting partners. Hishigaki, *et al.* (2001) proposed the chi–square method for n-neighbor protein interactions. Deng, *et al.*(2003) developed a Markov random field (MRF) method for protein function prediction using multiple PPI data. Lanckriet, *et al.* (2004) proposed a diffusion kernel based support vector machine (SVM) approach for predicting protein functions on a protein interaction network with 5-types data, the prediction accuracy of the SVM approach is higher than that ofthe MRF approach. Lee, *et al.*(2006) developed a diffusion kernel logistic regression (DKLR) method for protein interaction networks, which incorporates all neighbors of proteins in the network. The result showed that the DKLR approach of incorporating all protein neighbors significantly improved the accuracy of protein function predictions over the MRF model. And the prediction accuracy was comparable to another protein function classifier based on SVM with a diffusion kernel. Gao *et al.* (2007) developed a method to PPI data by introducing several methods to filter the neighbors in protein interaction networks for a protein of unknown functions. However, DKLR method can be expressed as a features extraction method and be calculated with traditional LR classifier for protein interaction prediction, and it has the different learning algorithm with the traditional KLR (Zhu, *et al.* 2002; Cawlay, *et al.* 2005; Birkenes, , *et al.* 2007; Tenenhaus, *et al.* 2007; Jaakkola, *et al.* 1999; Keerthi, *et al.* 2005; Karmakers, *et al.* 2007; Muller , *et al.* 2001 ).

From the pattern recognition point of view, the description of PPI information, namely PPI feature extraction and selection, takes crucial role in bio-informatics analysis (Saeys *et al.* 2007; Liu *et al.* 2009 ). To improve the prediction accuracy, we propose a 2-order graphic neighbor information extraction method for PPI prediction, which is different from Lee, *et al.*(2006) . We investigate the prediction performances of four classifiers, LR, DKLR, PKLR and RBF KLR on the 1-order graphic neighbor feature sets and 2-order graphic neighbor feature sets. Furthermore, the chi-square test method is also employed for multiple category feature combination. Since the first top chi-square value of one function may not be itself, we propose a new L-top chi-square feature combination method in section 4.3.3 and section 4.3.4 to overcome the insufficiency, that the own function feature vector should be list No.1and the feature vectors calculated by L-top chi-square values are attached behind to construct the new combination feature. The experimental results of the average overall percentage criterion on MIPS data ( Lu, *et al.* 2006; Ashburner, *et al.* 2000; Mewes, *et al.* 2002 ) of YPD demonstrate the effectiveness of 2-order neighbor graphic information and the performance of RBF KLR model.

The rest of the paper is organized as follows. Logistic regression and Kernel Logistic regression for a single function are reviewed in Section 2., and the diffusion kernel logistic model is also reviewed in Section 2. Kernel logistic regression with steepest descent Newton-Raphson method, 2-order graphic neighbor feature extraction method and feature combination with chi-square test are proposed and discussed in Section 3. The experimental results on MIPS database of YPD are given in Section 4. Finally, the conclusion is given in Section 5.

# 2. Review of Logistic regression and Kernel logistic regression

## *2.1. Logistic regression*

Logistic regression model is a popular statistical model (Hastie *et al.* 2009). Assume that $N$ independent and identically distributed (i.i.d.) samples $\{(x_i, y_i)\}_{i=1}^{N}$, $x_i \in R^d$, $y_i \in \{0,1\}$, satisfy unknown joint probability distribution $P(x, y)$. We denote $\Omega = \{x_1, \ldots, x_N\}$.

In the training space $\Omega$, the discriminative function of the two–class is defined as

$$g(x) = \ln(\frac{P(y=1|x)}{1-P(y=1|x)}) = \ln(\frac{P(y=1|x)}{P(y=0|x)}). \tag{1}$$

and the linear function $g_w(x) = w_0^T x + b$ is applied to fit $g(x)$. Let $x^* = [1 x^T]^T$, $w = [b w_0^T]^T$, we rewrite $g_w$ as $g_w(x) = w^T x^*$, and reemploy the symbols of $x$ and $x_i$ to denote $x^*$ and $x_i^*$ respectively.

Given sample $x$, the posterior probability of ``$y=1|x$'' is denoted as:

$$\pi_w(x) = P(y=1|x) = \frac{1}{1+\exp(-w^T x)}. \tag{2}$$

The parameter $w$ is calculated by maximizing conditional log–likelihood,

$$\begin{aligned} l(w) &= \sum_{i=1}^{N} \{y_i \ln(\pi_w(x_i)) + (1-y_i)\ln(1-\pi_w(x_i))\} \\ &= \sum_{i=1}^{N} \{y_i(w^T x) - \ln(1+\exp(w^T x))\}. \end{aligned} \tag{3}$$

The updating estimation of parameter with Newton-Raphson method is as follows,

$$\begin{aligned} w^{new} &= w^{old} - (\frac{\partial^2 l(w)}{\partial w \partial w^T})^{-1} \frac{\partial l(w)}{\partial w} = w^{old} - (X^T W X)^{-1} X^T (y-p) \\ &= (X^T W X)^{-1} X^T W (X w^{old} + W^{-1}(y-p)) = (X^T W X)^{-1} X^T W z \end{aligned} \tag{4}$$

where $y = [y_1, \ldots, y_N]^T$, $X = [x_1, \ldots, x_N]^T$, $p = [\pi(x_1, w^{old}), \ldots, \pi(x_N, w^{old})]^T$, $W = diag\{\pi(x_1, w^{old})(1-\pi(x_1, w^{old})), \ldots, \pi(x_N, w^{old})(1-\pi(x_N, w^{old}))\}$, $z = X w^{old} + W^{-1}(y-p)$.

Adding a penalty regularization item to $l(w)$, we obtain the ridge regression model (Roth, 2001)

$$H(w) = -\sum_{i=1}^{N} \{y_i(w^T x) - \ln(1+\exp(w^T x))\} + \frac{\lambda}{2} \|w\|^2. \tag{5}$$

where $\lambda$ balances the regression function and loss function. And the parameter $w$ is obtained by minimizing $H(w)$.

$$\begin{aligned} w^{new} &= w^{old} - (\frac{\partial^2 H(w)}{\partial w \partial w^T})^{-1} \frac{\partial H(w)}{\partial w} \\ &= (X^T W X + \lambda I)^{-1} X^T W z \end{aligned} \tag{6}$$

where $y = [y_1, \ldots, y_N]^T$, $X = [x_1, \ldots, x_N]^T$, $p = [\pi(x_1, w^{old}), \ldots, \pi(x_N, w^{old})]^T$, $W = diag\{\pi(x_1, w^{old})(1-\pi(x_1, w^{old})), \ldots, \pi(x_N, w^{old})(1-\pi(x_N, w^{old}))\}$, $z = X w^{old} + W^{-1}(y-p)$,

and $I$ is the identity matrix.

## 2.2. Kernel logistic regression model

Kernel logistic regression uses kernel trick to generalize logistic regression to high dimension feature space. Suppose $\kappa(x,y)$ is any kernel function satisfying Mercer condition, $\kappa$ defines a non–linear mapping from sample space to reproducing kernel Hilbert space （RKHS） space $\Phi: x \to \Phi(x)$.

Since the optimal vector which satisfies (3) is in the space of sample $x_i$,

$$w = \sum_{i=1}^{N} \alpha_i x_i. \tag{7}$$

we define the discriminative function in the RKHS space as,

$$g_w(x) = w^T \Phi(x) \tag{8}$$

As the parameter $w$ is also in the space of sample $\Phi(x_i)$, we can get

$$w = \sum_{i=1}^{N} \alpha_i \Phi(x_i). \tag{9}$$

Hence,

$$g_\alpha(x) = g_w(x) = (\sum_{i=1}^{N} \alpha_i \Phi(x_i))\Phi(x) = \sum_{i=1}^{N} \alpha_i \kappa(x_i, x). \tag{10}$$

And, the estimation of parameter $w$ turns to the estimation of parameter $\alpha$. Denote the posterior probability,

$$\pi_\alpha(x) = \frac{1}{1 + \exp(-\sum_{i=1}^{N} \alpha_i \kappa(x_i, x))}. \tag{11}$$

Similar to formula (5), the objective function of the ridge kernel logistic regression is defined as,

$$\begin{aligned} H(\alpha) &= -\sum_{i=1}^{N} \{y_i g_\alpha(x_i) - \ln(1 + \exp\{g_\alpha(x_i)\})\} + \frac{\lambda}{2} \| g_\alpha \|^2 \\ &= -\sum_{i=1}^{N} \{y_i K_i \alpha - \ln(1 + \exp\{K_i \alpha\})\} + \frac{\lambda}{2} \alpha^T K \alpha. \end{aligned} \tag{12}$$

where $K = [\kappa(x_i, x_j)]_{N \times N}$ is kernel matrix, $K_i$ is the $i$–th line of kernel matrix K.

The update estimation of $\alpha$ using Newton Raphson method is to find a vector to minimize the formula (12),

$$\alpha^{new} = (K_{N \times N}^T W K_{N \times N} + \lambda K_{N \times N})^{-1} K_{N \times N}^T W \tilde{z} \tag{13}$$

where $\tilde{z} = (K\alpha^{old} + W^{-1}(y - p))$.

A sparse solution of $\alpha$ is only involving a subset $S$ of training data as the basic functions, (Katz, *et al.* 2005 )

$$g_\alpha(x) = \sum_{i=1}^{s} \alpha_i K(x_i, x), s \ll N. \tag{14}$$

The ridge kernel logistic regression is defined as

$$H(\alpha) = -\sum_{i=1}^{s} \{y_i K_i \alpha - \ln(1 + \exp\{K_i \alpha\})\} + \frac{\lambda}{2} \alpha^T K_{s \times s} \alpha. \tag{15}$$

We obtain the sparse updating formula as follows,
$$\alpha^{new} = (K_{N\times s}^T W K_{N\times s} + \lambda K_{s\times s})^{-1} K_{N\times s}^T W \tilde{z} \qquad (16)$$
where $\tilde{z} = (K_{N\times s}\alpha^{old} + W^{-1}(y-p))$, $K_{N\times s} = [\kappa(x_i, x_j)]_{N\times s}$, $x_i \in \Omega, x_j \in S$, and, $K_{s\times s} = [\kappa(x_i, x_j)]_{s\times s}$, $x_i, x_j \in S$.

However, especially in PPI analysis, the high values of $s$ or $N$ make the matrix be irreversible, and the improper selection of $s$ vector of sample feature $x_i$ also leads to bad convergence of KLR. In this paper, we adopt the steepest descend Newton Raphson method (Karsmakers, *et al.* 2007) to estimate parameter $\alpha$ forKLR, it is a sub-optimization strategy compared to the formula (13) and (16). The steepest descend method based updating algorithm is as follows,
$$\alpha^{new} = \alpha^{old} - \delta \frac{\partial H(\alpha)}{\partial \alpha} \qquad (17)$$
where $\delta > 0$ is the step factor. Though the steepest descend method may not reach the global optimization value, it will avoid the large matrix reversible calculation, and accelerate the computation speed.

The widely used kernel functions are RBF kernel function and Polynomial function, where RBF kernel function is,
$$\kappa(x_i, x_j) = e^{-\gamma \|x_i - x_j\|_2^2}, \gamma > 0. \qquad (18)$$
And, Polynomial function is：
$$\kappa(x_i, x_j) = (x_i \cdot x_j + b)^h, b > 0, h > 0. \qquad (19)$$

## 2.3. Review of LR and KLR in PPI with 1-order graphic neighbor information

To model the interaction of proteins, Lee, *et al.* (2006) proposed a 1-order neighbor information feature extraction method and involved the logistic regression for PPI prediction for a given function,
$$\log(\frac{Pr(X_i = 1 | X_{[-i]}, \theta)}{1 - Pr(X_i = 1 | X_{[-i]}, \theta)}) = \gamma + \delta M_0(i) + \eta M_1(i). \qquad (20)$$
where,
$$X_{[-i]} = (X_1, \cdots, X_{i-1}, X_{i+1}, \cdots, X_N).$$
$$X_i = \begin{cases} 1, & \text{If protein i has the function.} \\ 0, & \text{If protein i does not have the function.} \end{cases}$$
$$M_0(i) = \sum_{j \neq i, x_j known} K(i,j) I_{\{x_j = 0\}}.$$
$$M_1(i) = \sum_{j \neq i, x_j known} K(i,j) I_{\{x_j = 1\}}. \qquad (21)$$
$$K(i,j) = \begin{cases} 1, & \text{If protein i interacts with protein j.} \\ 0, & \text{If protein i doesnot interact with protein j.} \end{cases}$$

And, $M_0(i)$ is equal to the number of proteins interacting with protein i, but without the function. $M_1(i)$ is equal to the number of proteins interacting with protein i and with the function. We call them 1–order graphic neighbor information, since they describe the direct interaction of proteins with its neighbors according to the graph theory.

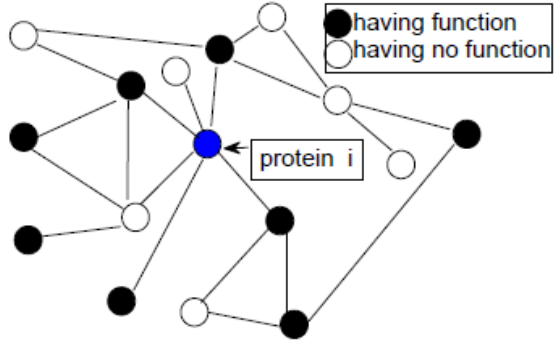

Figure 1. Protein–Protein interaction for one function

To generalize the Logistic regression, Lee et al (2006) extended the kernel $K(i,j)$ to the the diffusion kernel $K$, which calculates the similarity distance between any two nodes in the network. It is defined as follows:

$$K = \exp\{\tau H\}. \tag{22}$$

where,

$$H(i,j) = \begin{cases} 1 & \text{If protein i interacts with protein j.} \\ -d_i & \text{If protein i doesnot interact with protein j.} \\ 0 & \text{otherwise.} \end{cases} \tag{23}$$

and $d_i$ is the number of interaction partners for protein i, $\tau$ is diffusion constant, and $\exp\{H\}$ represents the matrix exponential of the adjacent matrix H.

The diffusion kernel logistic regression applies the formula (23)(22) to (21)(20), it is equal to a different feature extraction method for $M_0(i)$ and $M_1(i)$. From the theoretical point of view, it is a weighted graph modelling along the 1-order adjacent path.

To pursue high recognition accuracy, the chi-square test is employed to identify correlated functions for a function of interest in multi-function analysis. For a protein $P_i$ having a function $C_j$, the chi-square association value between the function $C_j$ and any function $C_i$ based on $P_i$'s immediate neighbors, is defined as

$$\frac{(N_i^{(1)}(l) - N_i^{(1)}Q_l)^2}{N_i^{(1)}Q_l}. \tag{24}$$

where $N_i^{(1)}$ is the number of immediate neighbors of $P_i$, $N_i^{(1)}(l)$ is the immediate neighbors of $P_i$ having function $C_i$, $Q_i$ is the fraction of known proteins having function $C_i$. Summing the corresponding quantities up over all proteins having function $C_i$ in the PPI network obtains an overall statistic

$$\frac{(\sum N_i^{(1)}(l) - \sum N_i^{(1)}Q_l)^2}{\sum N_i^{(1)}Q_l}. \tag{25}$$

When the first L-top chi-square values are considered, the chi–square logistic regression model is defined as follows,

$$\log(\frac{Pr(X_i = 1 | X_{[-i]}, \theta)}{1 - Pr(X_i = 1 | X_{[-i]}, \theta)}) = \alpha_0 + \sum_{j=1}^{L} \{\alpha_{1j} M_{0j}(i) + \alpha_{2j} M_{1j}(i)\} \tag{26}$$

where, $M_{0j}(i)$ is the number of immediate neighbors of protein $P_i$ in $j-$th top chi-square value function which have not the function. $M_{1j}(i)$ is the number of direct neighbors of protein $P_i$ in $j-$th top chi-square value function which have the function. In Lee *et al*(2006), $L=5$ is recommended for the experiment. This model would be applied to both logistic regression and diffusion kernel logistic regression.
Therefore, logistic regression, diffusion kernel logistic regression, and their corresponding chi–square regression models are logistic regression with different feature extractions in nature.

## 3. Kernel logistic regression and 2-order neighbor graph information

### 3.1. Kernel logistic regression

As analyzed above, the traditional KLR is different from the DKLR model proposed by Lee, *et al.* (2006). In our experiments, we will treat DKLR as a feature extraction method with $\tau$, and evaluate the traditional KLR with formula (17) of steepest descend Newton Raphson method on all the 1-order graphic features aforementioned in the section 2. The RBF kernel and Polynomial kernel will be applied in the experiments. And, the DKLR feature can also be calculated with traditional KLR.

### 3.2. 2-order graphic neighbor information

As shown in Fig 1, the 1-order graphic neighbor information describes the direct interaction protein information according to whether it has the function, the proteins in the interaction network which have indirect interaction with the target protein, also take effects to the function prediction. For the target protein i, we denote $\Omega_i$ as the 1–order neighbor set of protein i, where

$$\Omega_i = \{x_j \mid x_j \in \Omega, x_j \text{ interacts with } x_i, j \neq i\}, (1 \leq i \leq N).$$

And a novel 2-order information in the protein interaction network is proposed as follows,

$$\log(\frac{Pr(X_i = 1 \mid X_{[-i]}, \theta)}{1 - Pr(X_i = 1 \mid X_{[-i]}, \theta)}) = \gamma + \delta M_0(i) + \eta M_1(i) + \xi M_2(i) + \eta M_3(i). \quad (27)$$

where

$$X_{[-i]} = (X_1, \ldots, X_{i-1}, X_{i+1}, \ldots, X_N).$$

$$X_i = \begin{cases} 1, & \text{If protein i has the function.} \\ 0, & \text{If protein i does not have the function.} \end{cases}$$

$$M_0(i) = \sum_{j \neq i, x_j known} K(i,j) I_{\{x_j = 0\}}.$$

$$M_1(i) = \sum_{j \neq i, x_j known} K(i,j) I_{\{x_j = 1\}}.$$

$$M_2(i) = \sum_{x_j \in \Omega_i} \sum_{x_k \Omega_i, k \neq j, k \neq i, x_k known} K(j,k) I_{\{x_k = 0\}}.$$

$$M_3(i) = \sum_{x_j \in \Omega_i} \sum_{x_k \Omega_i, k \neq j, k \neq i, x_k known} K(j,k) I_{\{x_k = 1\}}.$$

$$K(i,j) = \begin{cases} 1, & \text{If protein i interacts with protein j.} \\ 0, & \text{If protein i doesnot interact with protein j.} \end{cases} \quad (28)$$

Note that,
1) For the feature of $[1, M_0(i), M_1(i), M_2(i), M_3(i)]^T, (1 \le i \le N)$, formula(27) and (28) give a 2–order neighbor logistic regression. If the K is substituted by diffusion kernel, the regression model will change to DKLR model. It's corresponding feature is denoted as

$$[1, M_0^\tau(i), M_1^\tau(i), M_2^\tau(i), M_3^\tau(i)]^T, 1 \le i \le N, \quad (29)$$

2) If the feature of $[1, M_0(i), M_1(i), M_2(i), M_3(i)]^T, (1 \le i \le N)$ is applied to KLR model, we will get 2-order graphic neighbor KLR. When the DKLR feature $[1, M_0^\tau(i), M_1^\tau(i), M_2^\tau(i), M_3^{tau}(i)]^T$ is taken into KLR, we also obtain a DKLR feature based KLR model.

3) All of the features of $[1, M_0(i), M_1(i), M_2(i), M_3(i)]^T (1 \le i \le N)$ and $[1, M_0^\tau(i), M_1^\tau(i), M_2^\tau(i), M_3^\tau(i)]^T (1 \le i \le N)$ can be applied to form corresponding L-top chi-squarefeature, and we will get 2-order graphic neighbor LR, DKLR and KLR models respectively.

4) By adding definitions, $M_{2j}(i)$ is the 2-order neighbors number of $P_i$ in $j-$th top chi-square function which have not the function; $M_{3j}(i)$ is the 2-order neighbors number of $P_i$ in $j-$th top chi-square function which have the function. $(j = 1,...,L)$, the top L chi-square values model of (26) will extended to 2-order top L chi-square logistic regression,

$$\log(\frac{Pr(X_i = 1 | X_{[-i]}, \theta)}{1 - Pr(X_i = 1 | X_{[-i]}, \theta)}) = \alpha_0 + \sum_{j=1}^{L} \{\alpha_{0j} M_{0j}(i) + \alpha_{1j} M_{1j}(i) + \alpha_{2j} M_{2j}(i) + \alpha_{3j} M_{3j}(i)\} \quad (30)$$

Also, we can get the corresponding chi-square based 2-order graphic neighbor LR, DKLR and KLR models.

From the above analysis, we can constitute different models of LR, DKLR and KLR according to the feature properties. At last, we give the detail updating algorithms of LR, DKLR and KLR involved in the experiments.

For a given feature $\{x_i, y_i\}_{i=1}^{N}$ in terms of formula (20)(22)(26)(27)(29)(30), we can train the LR, DKLR and KLR models according to the following two algorithms,

**Algorithm 1. Logistic regression**
1) Set maximum update times $T = 100$, ridge $\lambda > 0$, $\varepsilon = 1.0e-3$. Initialize $w^{(0)} = [0,.....,0]^T$, $t = 1$.
2) Calculate $w^{(1)}$ using formula (6) and calculate $H(w^{(1)})$ using formula(5).
3) For $t = t+1$, Calculate $w^{(t+1)}$ using formula (6) and calculate $H(w^{(t+1)})$ using formula(5).
4) If $|H(w^{(t+1)}) - H(w^{(t)})| < \varepsilon$ (Equally, $\|w^{(t+1)} - w^{(t)}\| < \varepsilon$.), or $t > T$, stop updating. Or, go to step 3) and repeat until convergence.

**Algorithm 2. Kernel logistic regression with steepest descent method**
1) Set maximum update times $T = 100$, ridge $\lambda > 0$, $\varepsilon = 1.0e-3$, select the kernel function (RBF or Polynomial). Initialize $\alpha^{(0)} = [0,.....,0]^T$, $t = 1$.
2) Calculate $\alpha^{(1)}$ using formula (17) and calculate $H(\alpha^{(1)})$ using formula(12).

3) For $t = t+1$, Calculate $\alpha^{(t+1)}$ using formula (17) and calculate $H(\alpha^{(t+1)})$ using formula (12).

4) If $|H(\alpha^{(t+1)}) - H(\alpha^{(t)})| < \varepsilon$ (Equally, $\|\alpha^{(t+1)} - \alpha^{(t)}\| < \varepsilon$.), or $t > T$, stop updating. Or, go to step 3) and repeat until convergence.

## 4. Experimental Results

### 4.1. Datasets and feature extraction

To predict the PPI function, All of the LR, DKLR and KLR models with corresponding features (including chi–square combination features) are applied to infer the protein function of yeast cellar from Yeast Proteome database (YPD, http://www.incyte.com/) and the PPI data from the Munich Information Center for Protein Sequences (MIPS, http://mips.gfs.de/). To obtain a reliable experimental evaluation result, we adopt the database evaluated in Deng, *et al.* . Both "YPD function category–cellular role" and "MIPS Physical interactions" data files are downloaded from *http://www.cmb.usc.edu /msms/FunctionPrediction/*. There are 43 known cellular functions ( including "other ") in YPD, and 2559 MIPS interaction pairs with names in YPD.

To evaluate the PPI function performances according to undirected graph, we first delete the self–interaction data and the symmetric–interaction data in MIPS data, which means that the ``$A - A$'' type data will not appear in the feature extraction procedure and the ``$A - B$'', ``$B - A$'' type data will be calculate once. The proteins on the final MIPS data can fit a simple undirected graph–no loop and no multiple edges (as Fig 1.). Finally, there are 43 function categories and each function has 1282 different proteins in our sample space.

Then, each protein's 1-order and 2-order graphic neighbor features are calculated according to the MIPS data , and its function is labeled according to the YPD function tabular. The feature of $[1 M_0(i) M_1(i) M_2(i) M_3(i)]$ represents the 1-order graphic neighbor information $\{M_0(i), M_1(i)\}$ and 2–order graphic neighbor information $\{M_2(i), M_3(i)\}$ simultaneously. Considered the diffusion kernel logistic regression, we will obtain another feature $[1 M_0^\tau(i) M_1^\tau(i) M_2^\tau(i) M_3^\tau(i)]$ with different $\tau$ for DKLR.

Then, for a fixed protein function, the combination features with L-top chi-square values construct the different protein prediction tasks. We list them in table 1,

Table 1: Feature sets of LR,DKLR and KLR

| Feature set | Feature vector combination | Notation |
|---|---|---|
| $F_1$ | $[1 M_0(i) M_1(i)]$ | 1–order feature |
| $F_2$ | $[1 M_0(i) M_1(i) M_2(i) M_3(i)]$ | 2–order feature |
| $F_3$ | $[1 M_0^\tau(i) M_1^\tau(i)]$ | diffusion 1–order feature |
| $F_4$ | $[1 M_0^\tau(i) M_1^\tau(i) M_2^\tau(i) M_3^\tau(i)]$ | diffusion 2–order feature |
| $F_5$ | $[1 \{M_{0k}(i) M_{1k}(i)\}_{k=1}^{L}]$ | L-chi-square of 1–order feature |
| $F_6$ | $[1 \{M_{0k}(i) M_{1k}(i) M_{2k}(i) M_{3k}(i)\}_{k=1}^{L}]$ | L-chi-square of 2–order feature |
| $F_7$ | $[1 \{M_{0k}^\tau(i) M_{1k}^\tau(i)\}_{k=1}^{L}]$ | diffusion L-chi-square of |

| | | 1–order feature |
|---|---|---|
| $F_8$ | $[1\{M_{0k}^{\tau}(i)M_{1k}^{\tau}(i)M_{2k}^{\tau}(i)M_{3k}^{\tau}(i)\}_{k=1}^{L}]$ | diffusion L-chi-square of 2–order feature |
| $F_9$ | $[1\{M_{0k}(i)M_{1k}(i)M_{2k}(i)\}_{k=1}^{L}]$ | L-chi-square of 2–order feature |
| $F_{10}$ | $[1\{M_{0k}^{\tau}(i)M_{1k}^{\tau}(i)M_{2k}^{\tau}(i)\}_{k=1}^{L}]$ | diffusion L-chi-square of 2–order feature |

## 4.2. Criterion of protein–protein interaction function accuracy

To evaluate the various regression models, two popular criteria are wildly used in the PPI prediction , average overall percentage and sensitivity.

### 4.2.1. Average overall percentage

The traditional criterion is the correct prediction accuracy according to the following probability, If $Pr(X_i = 1) \geq 0.5$, we predict that protein i has the function. If $Pr(X_i = 1) < 0.5$, we predict that protein i doesn't have the function.

Average overall percentage is defined as the average prediction accuracy of all classes of proteins. Obviously, average overall percentage is the correct recognition rate in pattern recognition.

### 4.2.2. Sensitivity (SN) and False-positive

Another statistical criterion is the true positive, true negative, false positive, and false negative to compare the prediction accuracy for different methods, which are given in the following table.

Table 2: Statistical criteria

| | Predicted positive | Predicted negative |
|---|---|---|
| Real positive | True positive, TP | False positive, FP |
| Real negative | False positive, FP | True negative, TN |

The standard performance measures for the classification problem based on these four values are sensitivity (SN) and false-positive (FPR) defined as follows:

$$SN = \frac{TP}{TP + FN}, FPR = \frac{FP}{TN + FP}. \qquad (31)$$

Since, correct recognition rate is the popular standard in pattern recognition, we adopt average overall percentage as criterion in our experiments.

## 4.3. Performance evaluation of LR, DKLR and KLR

Generally, all of the parameters in the LR, DKLR and KLR should be evaluated by cross–validation or ROC curve, for example, $\lambda$, $\tau$, *etc.* . Since our purpose is to evaluate the novel features and logistic regression models, we select the parameters with LR as benchmark, then apply the parameters to the RBF KLR.

### 4.3.1. Comparison of performances of LR and KLR on 1-order and 2-order feature sets in one function

Firstly, we investigate the ridge of $\lambda \in \{0.00001, 0.0001, 0.001, 0.01, 0.1, 1, 10, 100,$

1000,10000,100000} (namely, $\log_{10}\gamma \in \{-5,-4,-3,-2,-1,0,1,2,3,4,5\}$) for LR and set the diffusion parameter of DKLR $\tau \in \{0.00001, 0.0001, 0.001, 0.01, 0.1\}$ for 1–order feature and 2–order feature set respectively. We choose the parameters of $(b,h) \in \{0,0.1\} \times \{\frac{1}{2},1\}$ for PKLR, and set $\gamma \in \{0.0001, 0.001, 0.01, 0.1, 1, 10, 100, 1000\}$ for RBF KLR.

Theoretically, the parameter of ridge could be determined by cross–validation, we adopt the strategy that we examine the ridge and other parameters on feature sets $F_1$, $F_2$, $F_3$ and $F_4$, then we apply the optimized parameters for the further experiments. The average overall percentage values of LR and DKLR on 1–order and 2–order features with **Algorithm 1** are shown in Fig.2 and Fig.3.

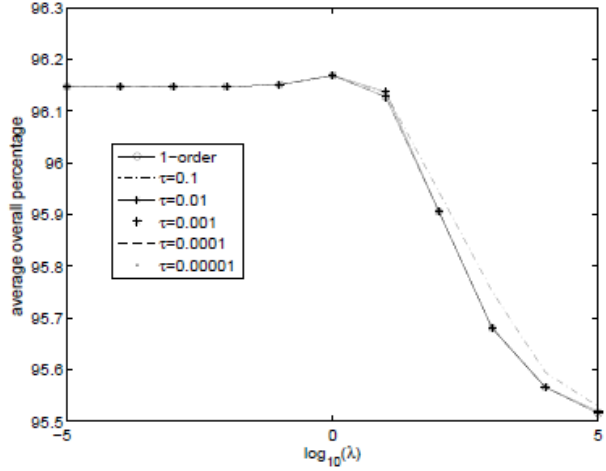

**Figure 2. 1–order graphic neighbor feature with LR and DKLR**

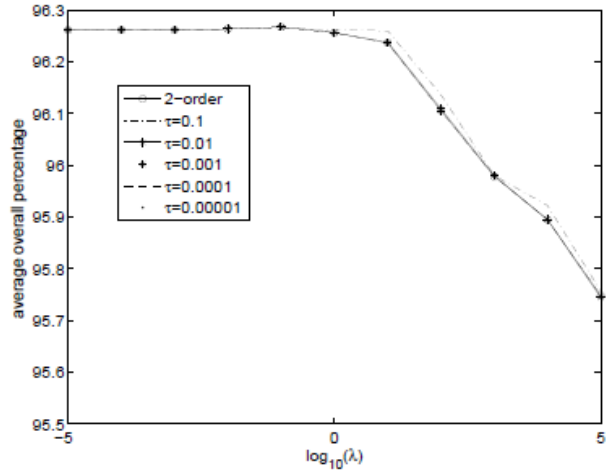

**Figure 3. 2–order graphic neighbor feature with LR and DKLR**

From Fig.2 and Fig3., we can conclude that 2-order graphic neighbor feature outperforms the corresponding 1-order graphic neighbor feature with both LR and DKLR models. As for the DKLR model, the diffusion kernel features $F_3$ and $F_4$ with the diffusion parameter $\tau = 0.1$ outperform the features $F_1$ and $F_2$ in traditional LR model. At the same time, the ridge $\lambda = 0.00001$ is a reasonable choice both for LR and DKLR.

Secondly, we compare the performance of LR, DKLR(with $\tau = 0.1$), PKLR and RBF

KLR on feature sets $F_1, F_2, F_3$ and $F_4$ respectively.

We only investigate the parameters of $(b,h) \in \{0, 0.1\} \times \{\frac{1}{2}, 1\}$ for PKLR, and the parameters of $\gamma \in \{0.001, 0.01, 0.1, 1, 10, 100, 1000\}$ for RBF KLR. The KLR models are trained with **Algorithm 2**. For the performances of PKLR trained by **Algorithm 2** with parameters of $(b,h) \in \{0, 0.1\} \times \{\frac{1}{2}, 1\}$ on $F_1, F_2, F_3$ and $F_4$, we only report the best results on the same feature set. For the average overall percentages of RBF KLR, we report the best results with $\gamma = 1000$.

Table 3: Average overall percentages(%) of 1-order and 2-order features with LR, DKLR(with $\tau = 0.1$), polynomial KLR and RBF KLR($\gamma = 1000$) with ridge $\lambda = 0.00001$

| Feature set | LR | PKLR | RBF KLR |
|---|---|---|---|
| $F_1$ | 96.15 | 93.34 | 96.60 |
| $F_2$ | 96.26 | 94.85 | 98.30 |
| $F_3, \tau = 0.1$ | 96.15 | 93.78 | 96.60 |
| $F_4, \tau = 0.1$ | 96.26 | 93.37 | 98.30 |

From Table 3, we can draw the conclusion that RBF KLR model with 1-order and 2-order graphic neighbor information outperforms LR and PKLR, and LR performs better than PKLR. Simultaneously, 2-order graphic neighbor information can improve the average overall percentage criterion compared to 1-order graphic neighbor information both with LR and RBF KLR.

### 4.3.2. Traditional L-top Chi-square feature combination with LR and KLR

We discuss the protein prediction rate on one function feature $F_1, F_2, F_3$ and $F_4$, and the above experimental results demonstrate the effectiveness of 2-order graphic neighbor feature. As discussed in the Lee *et. al.*(2006), we will demonstrate the effectiveness of 2-order graphic neighbor combination feature with chi-square method. The features of $F_1, F_2, F_3$ and $F_4$ will be extended tothe L-top chi-square base 1-order and 2-order feature sets respectively, denoted as $F_5$, $F_6$, $F_7$ and $F_8$. The average overall percentage values are list in Table 4.

Table 4: Average overall percentages(%) of chi-square based 1-order and 2-order combination features with LR, DKLR(with $\tau = 0.1$), and RBF KLR ($\gamma = 1000$) with ridge $\lambda = 0.00001$.

| Feature set | L | LR | RBF KLR |
|---|---|---|---|
| $F_5$ | 3 | 96.19 | 97.14 |
| | 4 | 96.20 | 97.29 |
| | 5 | 96.26 | 97.42 |
| $F_6$ | 3 | 96.38 | 98.77 |
| | 4 | 94.08 | 98.88 |
| | 5 | 94.10 | 98.97 |
| $F_7, \tau = 0.1$ | 3 | 96.19 | 97.14 |
| | 4 | 96.20 | 97.29 |
| | 5 | 96.27 | 97.42 |

| $F_8, \tau = 0.1$ | 3 | 96.39 | 98.77 |
| --- | --- | --- | --- |
| | 4 | 96.41 | 98.88 |
| | 5 | 94.10 | 98.97 |

### 4.3.3. A new L-top Chi-square feature combination with LR and KLR

Given one function, any protein's own feature's chi-square value may not be the maximum one, then the L-top chi–square features may not include its own feature, we have investigated the 1-order and 2-order chi–square feature's performances in Table 4. Since 1-order and 2-order features are the real data observed from protein interaction partners, it should included in the feature sets, we propose a revision of the chi–square feature combination that the first component vector is its own feature, and the L-top chi–square features are attached in succession. The average overall percentages with LR and RBF KLR are list in Table 5.

Table 5: Average overall percentages(%) of new L top chi-square based 1-order and 2-order combination features with LR, DKLR(with $\tau = 0.1$), and RBF KLR ($\gamma = 1000$) with ridge $\lambda = 0.00001$.

| Feature set | L | LR | RBF KLR |
| --- | --- | --- | --- |
| $F_5$ | 3 | 96.19 | 97.15 |
| | 4 | 96.26 | 97.29 |
| | 5 | 100.00 | 100.00 |
| $F_6$ | 3 | 96.39 | 98.86 |
| | 4 | 96.42 | 98.96 |
| | 5 | 96.43 | 99.05 |
| $F_7, \tau = 0.1$ | 3 | 96.19 | 97.14 |
| | 4 | 96.20 | 97.29 |
| | 5 | 100.00 | 100.00 |
| $F_8, \tau = 0.1$ | 3 | 96.40 | 98.86 |
| | 4 | 96.42 | 98.96 |
| | 5 | 94.11 | 99.05 |

It can be concluded that on the same combination feature set of both 1-order and 2-order neighbor features, RBF KLR model outperforms LR model. And, on the 5 top 1-order chi-square based combination feature and corresponding DKLR feature sets, LR and RBFKLR can achieve 100% right prediction accuracy. On 5-top 2-order chi-square based combination feature sets, RBF KLR can reaches 99.05% average overall percentage, though this kinds of feature may introduce more redundancy, as the dimension of L-top Chi-square 2-order feature vector ($F_6$) is almost two times of the dimension of L-top Chi-square 1-order feature vector ($F_5$),.

### 4.3.4. Another L-top Chi-square 2-order feature combination

As discussed in Table 5, it it shown that RBF KLR can achieve the average overall percentage beyond 99.05% with the new 5-top chi-square combination on all 1-order features ($F_5$, $F_7(\tau = 0.1)$) and 2-order features ($F_6$, $F_8(\tau = 0.1)$). we propose another new L-top chi-square 2-order feature combination method of $F_9$ and ($F_{10}$, $\tau = 0.1$), and the combination rule obeys the way in section 4.3.3, that the first combination vector should be its own and the vectors determined by L-top chi-square are attached sequentially. The prediction accuracy results of $F_9$ and ($F_{10}$, $\tau = 0.1$) are list in Table 6.

Table 6: Average overall percentages(%) of L top chi-square based 2-order combination features with LR, DKLR(with $\tau$ = 0.1), and RBF KLR ($\gamma$ = 1000 ) with ridge $\lambda$ =0.00001.

| Feature set | L | Traditional L-top chi-square | | New L-top chi-square | |
| --- | --- | --- | --- | --- | --- |
| | | LR | RBF KLR | LR | RBF KLR |
| $F_9$ | 3 | 96.32 | 98.75 | 96.33 | 98.84 |
| | 4 | 96.37 | 98.88 | 96.38 | 98.95 |
| | 5 | 96.39 | 98.97 | 96.40 | 99.05 |
| $F_{10}$, $\tau = 0.1$ | 3 | 96.32 | 98.75 | 96.33 | 98.84 |
| | 4 | 94.05 | 98.88 | 96.38 | 98.96 |
| | 5 | 94.07 | 98.97 | 96.40 | 99.05 |

The experimental results confirm that our new L-top 2-order neighbor combination feature provides more robustness information than traditional L-top 2-order neighbor combination feature, and on this kinds of feature selection method, RBF KLR still reaches 99.05% average overall percentage.

In addition, from Table 4, both LR and RBF KLR can achieve 100% average overall percentage, how to reveal the low dimension of new L-top 2-order neighbor combination feature will be our farther research problem. Since all of our models are developed and tested on the same known feature sets, there will be unavoidable overfitting problem and dimension calamity, the L-top chi-square 2-order feature combination in Section 4.3.4 is obviously a strategy of dimension reduction of L-top chi-square 2-order feature combination in Section 4.3.3.. At the same time, the experimental results demonstrate the robustness of 2-order graphic information in PPI prediction. Since we have achieved relative high average overall percentage, the sensitivity criterion is omitted in this paper.

# 5. Conclusion

A 2-order graphic neighbor information extraction method is proposed for PPI prediction, and the chi–square based feature combination is also involved to improve the prediction accuracy. To demonstrate its effectiveness in one function prediction, LR, DKLR, PKLR and RBF KLR are involved in protein function prediction. The experimental results show that RBF KLR can achieve high average overall percentage value for PPI especially with our two new 5-top chi–square based 2-order graphic neighbor combination features. The future work will focus on applying the graphic features and kernel logistic regression models to unknown protein function prediction and liver cancer microRNA network discovery .

**Acknowledgements**
This project was supported by 863 Project of China (2008AA02Z306). The author would like to thank Professor Minping Qian for valuable discussion, MS. Shuang Hu for typing some tables, MS. Kang Jin for data preparation, discussion and some analysis.

# References

[1]   B. Schwikowski, P. Uetz, S. Fields. A network of protein-protein interactions in yeast.

Nature America, 2000, 18: 1257–1261.
[2] H., Hishigaki, K., Nakai, T., Ono, A., Tanigami, T., Takagi. Assessment of prediction accuracy of protein function from protein–protein interaction data. Yeast, 2001 , 18: 523–531.
[3] M.H. Deng, K. Zhang, S. Mehta, T. Chen, F.Z. Sun. Prediction of protein function using protein-protein interaction data. Journal of Computational Biology. 2003, 10(6):947-60.
[4] G.R.G. Lanckriet, M.H. Deng, N. Cristianini, M.I. Jordan, W.S. Noble. Kernel-based data fusion and its application to protein function prediction in yeast. Proceedings of the Pacific Symposium on Biocomputing, January 3-8, 2004. pp. 300-311.
[5] H. Lee, Z. Tu, M.H. Deng, F.Z. Sun, Y. Chen. Diffusion kernel-based logistic models for protein function prediction. OMICS: A Journal of Integrative Biology. 2006, 10(1): 40-55.
[6] L. Gao, X. Li, Z. Guo, M.Z. Zhu, Y.H. Li, S.Q. Rao. Widely predicting specific protein functions based on protein-protein interaction data and gene expression profile. Science in China Series C: Life Sciences, 2007,50(1):125-134.
[7] V. Roth. Probabilistic discriminative kernel classifiers for multi-class problems. In: Radig B., Florczyk S. (eds.), Pattern Recognition–DAGM'01, p. 246-253, Springer, LNCS 2191.
[8] J. Zhu, T. Hastie. Support vector machines, kernel logistic regression and boosting. Computer Science. 2002, Vol. 2364,16-26.
[9] G.C. Cawley, N.L.C. Talbot. The evidence framework applied to sparse kernel logistic regression. Neurocomputing, 2005, 64: 119-135.
[10] $\varphi$. Birkenes, T. Matsui, K. Tanabe, T.A. Myrvoll. N-best rescoring for speech recognition using penalized logistic regression machines with garbage class. IEEE International Conference on Acoustics, Speech and Signal Processing, 2007, 4: IV-449-IV-452.
[11] A. Tenenhaus, A. Giron, E. Viennet, M. Bera, G. Saporta, B. Fertile. Kernel logistic PLS: A tool for supervised nonlinear dimensionality reduction and binary classification. Computational Statistics & Data Analysis, 2007, 51: 4083-4100.
[12] T.S. Jaakkola , D. Haussler. Probabilistic kernel regression models. In Proceedings of the 1999 Conference on AI and Statistics . Morgan Kaufmann, 1999.
[13] S.S. Keerthi, K.B. Duan, S.K. Shevade, A.N. Poo. A fast dual algorithm for kernel logistic regression. Machine Learning, 2005, 61(15): 151-165.
[14] P. Karsmakers, K. Pelckmans, J. Suykens, H.V. Hamme. Fixed-size kernel logistic regression for phoneme classification. In: 8th Annual Conference of the International Speech Communication Association (INTERSPEECH) 2007, 78-81.
[15] T. Hastie,R. Tibshirani,J. Friedman. The Elements of Statistical Learning: Data Mining, Inference, and Prediction. Second Edition. Springer, 2009.
[16] M. Katz, M. Schaffoner, E. Andelic, S. Kr$\ddot{u}$ ger, A. Wendemuth. Sparse Kernel Regression for Phoneme Classification. Proceedings of 10th International Conference on Speech and. Computer (SPECOM), 2005, October 05, Patras, Greece, vol.2 : 523-526.
[17] K.R. Muller, S. Mika, G. Ratsch, K. Tsuda, B. Scholkopf. An introduction to kernel-based learning algorithms. IEEE Transactions On Neural Networks, 2001, 12:181-201
[18] Y. Saeys, I. Inza, P. Larranaga. A review of feature selection techniques in bioinformatics. Bioinformatics, 2007, 23(19): 2507-2517.
[19] J.W. Liu, M.H. Deng, M.P. Qian. Feature-based causal structure discovery in protein and gene expression data with Bayesian network. Fifth International Conference on Natural Computation 2009, 144-148.
[20] P.T. Spellman, G. Sherlock, M.Q. Zhang, V.R. Iyer, K. Anders, M.B. Eisen, P.O. Brown, D. Botstein, B. Futcher. Comprehensive identification of cell cycle-regulated genes of the yeast saccharomyces cerevisiae by microarray hybridization. Molecular Biology of the Cell, 1998, 9: 3273 -3295.
[21] T.I. Lee, N.J. Rinaldi, F. Robert, D.T. Odom, Z.B. Joseph, G.K. Gerber, N.M. Hannett, C. T. Harbison, C.M. Thompson, I. Simon, J. Zeitlinger, E.G. Jennings, H.L. Murray, D.B. Gordon, B. Ren, J.J. Wyrick, J.B. Tagne, T.L. Volkert, E. Fraenkel, D.K. Gifford, R.A. Young. Transcriptional regulatory networks in sacchsromyces cerevisiae. Science, 2002, 298(5594): 799 - 804
[22] H.C. Lu, Q.Y. Shi, B.C. Shi, Z.H. Zhang, Y. Zhao, S.Q. Tang, L. Xiong, Q. Wang, R.S. Chen. Predicting Protein Function Based on Modularized Protein Interaction Network. Progress in Biochemistry and Biophysics, 2006, 33 (5): 446-451
[23] M. Ashburner, C.A. Ball, J.A. Blake, D. Botstein, H. Butler, J.M. Cherry, A.P. Davis, K.

Dolinski, S.S. Dwight, J.T. Eppig, M.A. Harris, D.P. Hill, L. Issel-Tarver, A. Kasarskis, S. Lewis, J.C. Matese, J.E. Richardson, M. Ringwald, G.M. Rubin, G. Sherlock. Gene ontology: tool for the unification of biology. Nature Genetics, 2000, 25: 25-29

[24] H.W. Mewes, D. Frishman, U. Guldener, G. Mannhaupt, K. Mayer, M. Morkrejs, B. Morgenstern, M. Munsterkotter, S. Rudd, B. Weil. MIPS: a database for genomes and protein sequences. Nucleic Acids Research, 2002, 30(1) : 31-34.

[25] R. Shalgi, D. Lieber, M. Oren, Y. Pilpel. Global and local architecture of the mammalian microRNA -transcription factor regulatory network. PLoS Computational Biology, 2007, 3(7): 1291- 1304.

[26] H. Varnholt. The role of microRNAs in primary liver cancer. Annals of Hepatology, 2008, 7(2): 104-113.

[27] C.W. Hu, H.F. Juan, H.C. Huang. Characterization of microRNA-regulated protein-protein interaction network . Proteomics, 2008, 8(10): 1975-1979.